\documentclass[
 aip,
 jmp,
 amsmath,amssymb,
reprint,%
]{revtex4-1}

\usepackage{graphicx}
\usepackage{dcolumn}
\usepackage{bm}
\usepackage[dvips]{color}

\begin{document}


\title{A note on the Duffin-Kemmer-Petiau equation in (1+1) space-time dimensions}

\author{Jos\'e T. Lunardi}
\email{jttlunardi@uepg.br}
\affiliation{School of Physics and Astronomy, University of Glasgow\\ G12 8QQ, Glasgow,UK}
\affiliation{Department of Mathematics \& Statistics, State University of Ponta Grossa\\ Avenida Carlos Cavalcanti 4748, Cep 84030-900, Ponta Grossa, PR, Brazil
}%

%


\begin{abstract}
In the last years several papers addressed the supposed spin-1 sector of the massive Duffin-Kemmer-Petiau (DKP) equation restricted to (1+1) space-time dimensions. In this note we show explicitly that this is a misleading approach, since the DKP algebra in (1+1) dimensions admits only a spin-0 representation. Our result also is useful to understand why several recent papers found coincident results for both spin-0 and spin-1 sectors of the  DKP theory in (3+1) dimensions when the dynamics is restricted to one space dimension.
\end{abstract}

\maketitle

%


The Duffin-Kemmer-Petiau (DKP) equation is a first order wave equation similar to the Dirac one, which in its original formulation in (3+1) space-time dimensions describes spin-0 and spin-1 fields or particles \cite{Pet36, Duf38, Kem38, Kem39, LPT00}. In recent years some papers addressed the DKP equation in strict (1+1) space-time dimensions in situations involving interactions and addressed the supposed spin-1 sector of the theory \cite{SHa06, Bou07, SHa10, DHS15}. In this note we show explicitly that such an approach is misleading; by using the (1+1)-dimensional analogs of the original DKP spin-0 and spin-1 projection operators we show that the supposed ``spin-1" sector of the theory restricted to (1+1) dimensions actually is unitarily equivalent to its spin-0 sector, which describes a (pseudo)scalar field. We illustrate this equivalence by explicitly building the lowest dimensional (irreducible) representation of the theory. At the end we comment how our result explain why several authors in recent years found identical results for both the spin-0 and spin-1 sectors of the (3+1) dimensional DKP equation when the dynamics is restricted to only one space dimension \cite{CMB04, BBo05, Bou08, CTC10, CCC10, Cas10, BBo12, CTr13, Bou15, dOl16, FMB17}.

\emph{DKP equation in (3+1) dimensions.} We start by recalling some basic results about the free DKP equation in (3+1) space-time dimensions. The equation is given by \cite{Pet36, Duf38, Kem38, Kem39, LPT00} (we use natural units $\hbar=c=1$)
\begin{equation}\label{dkp3}
\left(i\beta^\mu \partial_\mu - m\right)\psi=0\, \qquad \mu=0,1,2,3,
\end{equation}
where $m$ is the particle's mass, $\psi$ is the DKP wave function and $\beta^\mu$ are matrices satisfying the DKP algebra
\begin{equation}\label{dkpalgebra3}
\beta^\mu  \beta^\nu  \beta^\rho + \beta^\rho \beta^\nu \beta^\mu = g^{\mu\nu} \beta^\rho + g^{\rho\nu} \beta^\mu\, ,
\end{equation}
where $g^{\mu\nu}$ is the Minkowski metric tensor in (3+1) dimensions with signature $(+,-,-,-)$. It is well known that there are only three irreducible representations (irrep's) of DKP algebra in (3+1) dimensions: one is trivial, having dimension 1, and the other two are nontrivial, having dimensions 5 and 10, corresponding respectively to scalar (spin-0) and vector (spin-1) fields \cite{Kem39, Tok66, FNS73}.

Under infinitesimal Lorentz transformations $x^{\prime\mu}=\Lambda^{\mu}_{\nu}x^{\nu}$, with  $\Lambda^{\mu\nu}=g^{\mu\nu}+\omega^{\mu\nu}$, $\omega^{\mu\nu}=-\omega^{\nu\mu}$, the DKP spinor $\psi$ transforms as $\psi\to U\psi$, where \cite{LPT00}
\begin{equation}\label{LT}
U=1+\frac{1}{2}\omega^{\mu\nu}S_{\mu\nu},\qquad S_{\mu\nu}=\left[\beta_\mu,\beta_\nu\right]\, .
\end{equation}
The scalar and the vector sectors of the theory can be identified through the use of the Fujiwara's projectors \cite{Fuj53,LPT00}. For the scalar sector they are
\begin{eqnarray}\label{spin0proj}
P\!&=&\!-\left(\beta_0\right)^2\left(\beta_1\right)^2\left(\beta_2\right)^2\left(\beta_3\right)^2\\
\label{pmu3}
P^\mu\!&=&\!P\beta^\mu\, ,
\end{eqnarray}
which, from (\ref{dkpalgebra3}), satisfy
\begin{eqnarray}\label{lor3}
P\left(U\psi\right)\!&=&\!P\psi\\
P^{\mu}\left(U\psi\right)\!&=&\!P^{\mu}\psi\, .
\end{eqnarray}
These relations show that $P\psi$ transforms like a scalar and $P^{\mu}\psi$ transforms like a vector. By applying $P$ and $P^{\mu}$ on the DKP equation (\ref{dkp3}) we obtain the following relations
\begin{eqnarray}\label{pmu3div}
\partial_\mu \left(P^\mu\psi\right)\!&=&\!-im\left(P\psi\right)\\
\label{pmu3der}
P_\mu \psi \!&=&\!  \frac{i}{m}\partial_\mu \left(P\psi\right),
\end{eqnarray}
from which one concludes that the scalar $P\psi$ satisfy the Klein-Gordon equation $\left(\Box +m^2\right) P\psi=0$, with the elements of $P_\mu\psi$ being essentially the derivatives of the corresponding elements of $P\psi$.

Similarly, the vector sector is obtained by using the projectors \cite{Fuj53,LPT00}
\begin{eqnarray}\label{spin1pr13d}
R^\mu\!&=&\!\left(\beta_1\right)^2\left(\beta_2\right)^2\left(\beta_3\right)^2\left[\beta^\mu\beta^0-g^{\mu 0}\right]\\ \label{spin1pr23d}
R^{\mu\nu}\!&=&\!R^{\mu}\beta^{\nu}.
\end{eqnarray}
From (\ref{dkpalgebra3}) we conclude that $R^{\mu}\psi$ transforms as a vector, whereas $R^{\mu\nu}$ transforms as a second-rank antisymmetric tensor. Applying $R^{\mu}$ and $R^{\mu\nu}$ on the DKP equation (\ref{dkp3}) we obtain
\begin{eqnarray}\label{proca3d1}
&&\partial_\nu\left(R^{\mu\nu}\psi\right)= -im R^\mu\psi\\
\label{proca3d2}
&&R^{\mu\nu}\psi=-\frac{i}{m}U^{\mu\nu},\quad U^{\mu\nu}=\left[\partial^\mu\left(R^\nu\right)-\partial^\nu\left(R^\mu\right)\right],
\end{eqnarray}
which, combined, show that the field $ R^{\mu}\psi$ satisfy the Proca's equation
\begin{equation}
\left(\Box +m^2\right) R^{\mu}\psi=0, \qquad \partial_\mu\left(R^{\mu}\psi\right)=0, \,
\end{equation}
with $U^{\mu\nu}$ being merely the strength tensor.

From the above results one concludes that the operators $P$ and $P^\mu$ select the spin-0 sector whereas the operators $R^\mu$ and $R^{\mu\nu}$ select the spin-1 sector of DKP theory in (3+1) dimensions \cite{LPT00, Fuj53}. We recall that the product of one operator from the pair $\left(P,P^\mu\right)$ with any other operator from the pair $\left(R^\mu,R^{\mu\nu}\right)$ vanishes. This means that the spin-0 sector and the spin-1 sector are unequivalent irrep's of DKP algebra \cite{FNS73}. As mentioned above, the nontrivial irrep's of the DKP wave function $\psi$ for the spin-0 and spin-1 sectors correspond to ``spinors" having respectively 5 and 10 components. Explicit 5- and 10-dimensional irrep's for the matrices $\beta^\mu$ and the spinor $\psi$ can be easily obtained by rewriting respectively the Klein-Gordon and the Proca equations to a system of first-order differential equations. Another way to obtain an explicit (reducible) representation for the DKP matrices is from $\beta^\mu=\frac{1}{2}\left(\gamma^\mu\otimes I + I\otimes \gamma^\mu\right)$, where $\gamma^\mu$ are the $4\times 4$ Dirac matrices and $I$ is the $4\times 4$ identity. This $16\times 16$ representation is reducible to the already mentioned trivial (dimension 1), spin-0 (dimension 5) and spin-1 (dimension 10) irrep's \cite{Kem39}.

\emph{DKP equation in (1+1) dimensions.} To consider the DKP theory in (1+1) dimensions we now restrict the space-time labels in the equations (\ref{dkp3}) and (\ref{dkpalgebra3}) to $\mu,\nu=0,1$. The analogs of the spin-0 projectors (\ref{spin0proj})-(\ref{pmu3}) become
\begin{eqnarray}\label{p01}
P\!&=&\!-\left(\beta_0\right)^2\left(\beta_1\right)^2\\
\label{pmu1}
P^\mu\!&=&\!P\beta^\mu,\qquad \mu,\nu=0,1,
\end{eqnarray}
whereas the analogs in (1+1) dimensions of all the remaining equations (\ref{lor3})-(\ref{pmu3der}) do not change in form (only the space-time labels are restricted to the values $0,1$). In the same way we conclude that $P\psi$ transforms like a Lorentz scalar and $P^\mu\psi$ transforms like a vector, with $P\psi$ and $P^\mu\psi$ satisfying (\ref{pmu3div}) and (\ref{pmu3der}) (with $\mu,\nu=0,1$) and $P\psi$ satisfying the Klein-Gordon equation in (1+1) dimensions. So, as expected, the projectors $P$ and $P^\mu$ select the spin-0 (scalar) sector of the (1+1) dimensional DKP equation.

Now we address the main point of this note by considering the analogs of the ``spin-1" projectors (\ref{spin1pr13d})-(\ref{spin1pr23d}) in (1+1) dimensions:
\begin{eqnarray}\label{spin1pr11d}
R^\mu\!&=&\!\left(\beta_1\right)^2\left[\beta^\mu\beta^0-g^{\mu 0}\right]\\ \label{spin1pr21d}
R^{\mu\nu}\!&=&\!R^{\mu}\beta^{\nu},\qquad \mu,\nu=0,1.
\end{eqnarray}
It is straightforward to verify that $R^\mu\psi$ transforms like a vector under infinitesimal Lorentz transformations, but now we have that the only operators $R^{\mu\nu}$ which are non-vanishing in (1+1) dimensions are $R^{01}=-R^{10}$. Accordingly, under Lorentz transformations (\ref{LT}) ($\mu,\nu=0,1$)
\begin{equation}
R^{01}\left(U\psi\right)=R^{01}\psi\, ,
\end{equation}
since from DKP algebra we have that $R^{01}S_{01}=0$. Therefore, $R^{01}\psi$ transforms like a scalar (or like a pseudo-scalar, if we include improper Lorentz transformations). This is as expected, because in (1+1) dimensions any second-rank antisymmetric tensor must transform like a pseudo-scalar. Moreover, the ``spin-1" projectors (\ref{spin1pr11d})-(\ref{spin1pr21d}) can now be rewritten in terms of the spin-0 ones,
\begin{equation}\label{rels}
R^0=-\beta^1P^1,\quad R^1=-\beta^1 P^0, \quad R^{01}=\beta^1 P\, ,
\end{equation}
from which one can easily check that some of the products involving one operator from the pair $\left(P,P^\mu\right)$ and an operator from the pair $\left(R^\mu,R^{\mu\nu}\right)$ do not vanish. This is a consequence of the fact that the spin-0 and the ``spin-1" sectors of the (1+1) theory are no longer unequivalent irrep's of the algebra (\ref{dkpalgebra3}); indeed, they are unitarily equivalent, as we will show below. By using the above relationships the analogs of equations (\ref{proca3d1})-(\ref{proca3d2}) in (1+1) dimensions turn out to be
\begin{eqnarray}\label{vec1eq1}
&&\beta^1 \partial_1\left(P\psi\right)=\beta^1(im) P^1\psi \\
&&\beta^1 \partial_0\left(P\psi\right)=-\beta^1(im) P^0\psi \\
&&\beta^1\left(P\psi\right)=\beta^1\frac{i}{m}\left[\partial^0\left(P^0\psi\right)-\partial^1\left(P^1\psi\right)\right].
\label{vec1eq3}
\end{eqnarray}
By multiplying the above equations on the left by $\beta^1$, and taking into account that $-\left(\beta^1\right)^2P=P$, we obtain exactly the same set of equations (\ref{pmu3div})-(\ref{pmu3der}) with the space-time labels restricted to (1+1) dimensions. Conversely, if we multiply (\ref{pmu3div})-(\ref{pmu3der}) (with $\mu,\nu=0,1$) on the left by $\beta^1$ we obtain again (\ref{vec1eq1})-(\ref{vec1eq3}). Therefore, we conclude that the DKP spin-0 and ``spin-1" sectors are (unitarily) equivalent in (1+1) dimensions.

Summing up the above result, we conclude that \emph{there is no spin-1 sector in the strict (1+1)-dimensional DKP theory. The theory admits only a spin-0 sector, that corresponds to a (pseudo)scalar field}. Therefore, it is misleading to address a ``spin-1" representation of the DKP algebra in (1+1) space-time dimensions as a different representation beyond the spin-0 one, as it was done in the references \cite{SHa06, Bou07, SHa10, DHS15}.

In order to illustrate the above result we can easily build an explicit irrep for the DKP $\beta^\mu$ matrices ($\mu=0,1$) and ``spinor" $\psi$ by writing the (1+1)-dimensional second-order Klein-Gordon equation $\left(\Box+m^2\right)\phi=0$ for a scalar field $\phi$ as a system of first-order equations in the form (\ref{dkp3}). Doing so, we obtain the following $3\times 3$ irrep:
\begin{equation}\label{irrep1}
\beta^0\!=\!\left(
\begin{array}{ccc}
0&0&i\\
0&0&0\\
-i&0&0
\end{array}
 \right),\!\quad
\beta^1\!=\!\left(
\begin{array}{ccc}
0&i&0\\
i&0&0\\
0&0&0
\end{array}
 \right),\!\quad
\psi\!=\!\frac{1}{m}\!\left(
\begin{array}{c}
m\phi\\
\partial^1\phi\\
\partial^0\phi
\end{array}
\right).
\end{equation}
Similarly, we can write the (1+1)-dimensional Proca's equations
$$
\left(\Box +m^2\right)A^\mu=0, \quad \partial_\mu A^\mu=0, \quad \mu,\nu=0,1,
$$
as a system of first order equations in the DKP form and obtain exactly the same representation for the $\beta^\mu$ matrices as in (\ref{irrep1}), with $\psi$ now given by $\left(\frac{F^{01}}{m},-A^0,-A^1\right)^T$ ($T$ denotes the transpose), where $F^{01}=\partial^0 A^1-\partial^1 A^0$. From this result we can promptly map the fields in the two cases:
$$
\frac{F^{01}}{m}=\phi,\quad -A^0=\frac{1}{m}\partial^1\phi,\quad -A^1=\frac{1}{m}\partial^0\phi\, ,
$$
thus making explicit the equivalence between the spin-0 and ``spin-1" representations of the DKP theory in (1+1) dimensions.

Another way to obtain an explicit representation for the (1+1) dimensional DKP algebra is by the formula $\beta^\mu=\frac{1}{2}\left(\gamma^\mu\otimes I + I\otimes \gamma^\mu\right)$, where $\gamma^\mu$ ($\mu=0,1$) are now the $2\times 2$ Dirac matrices in (1+1) dimensions (which can be chosen as two Pauli matrices) and $I$ is the $2\times 2$ identity matrix. The authors of references \cite{SHa06, Bou07, SHa10, DHS15} claim that by using such representation they are addressing the spin-1 sector of the theory. This is not correct, since it is an easy task to show that this $4\times 4$ representation is in fact \emph{reducible} into a trivial representation of dimension 1 in which all $\beta^\mu=0$ ($\mu=0,1$) and $\psi=0$) and a nontrivial $3\times 3$  irrep that is unitarily equivalent to (\ref{irrep1}) which, as we had shown above, describes a spin-0 field. We also must mention a mistake which was propagated in these references, namely the lacking of the factor $1/2$ in the above expression for $\beta^\mu$ in terms of the Dirac matrices; without this factor the obtained $\beta^\mu$ matrices do not even fulfill the DKP algebra (\ref{dkpalgebra3}).

Our result is also useful to explain why in several recent papers the authors obtained the same results for both the spin-0 and spin-1 sectors of the (3+1) dimensional DKP theory when the dynamics was restricted to one space dimension \cite{CMB04, BBo05, Bou08, CTC10, CCC10, Cas10, BBo12, CTr13, Bou15, dOl16, FMB17}, a finding sometimes referred to as a remarkable one. We argue that this is not surprising; indeed, it is a trivial consequence of our main result, since the equation for the unidimensional propagation of the dynamic components of the wave function in the (3+1) case should be formally identical to a (1+1)-dimensional DKP equation, which, as we have shown here, admits only a spin-0 representation.

Summarizing, in this note we have shown in details that the DKP equation restricted to a (1+1)-dimensional space-time admits only a spin-0 [(pseudo)scalar] irreducible representation (appart from unitary transformations), which can be explicitly obtained in the form (\ref{irrep1}). This result shows that it is misleading to consider a genuine ``spin-1" (vector) sector of the theory. Our result also was useful to clarify the reason why the results for both the spin-0 and spin-1 sectors of the DKP theory in (3+1) dimensions coincide when the dynamics is restricted to only one space dimension.   

\emph{Acknowledgement.} The author thanks the anonymous referees for their comments and suggestions.

\bibliographystyle{unsrt}
\bibliography{NoteDKP1d}

\begin{thebibliography}{10}

\bibitem{Pet36}
G.~Petiau.
\newblock {\em Acad. R. Belg. Cl. Sci. Mem. Collect. 8}, 16, 1936.

\bibitem{Duf38}
R.~J. Duffin.
\newblock {\em Phys. Rev.}, 54:1114, 1938.

\bibitem{Kem38}
N.~Kemmer.
\newblock {\em Proc. R. Soc. A}, 166(924):127, 1938.

\bibitem{Kem39}
N.~Kemmer.
\newblock {\em Proc. R. Soc. A}, 173(952):91, 1939.

\bibitem{LPT00}
J.T. Lunardi, B.M. Pimentel, R.G. Teixeira, and J.S. Valverde.
\newblock {\em Phys. Lett. A}, 268(3):165, 2000.

\bibitem{SHa06}
K.~Sogut and A.~Havare.
\newblock {\em Class. Quantum Grav.}, 23(23):7129, 2006.

\bibitem{Bou07}
A.~Boumali.
\newblock {\em Phys. Scripta}, 76(6):669, 2007.

\bibitem{SHa10}
K.~Sogut and A.~Havare.
\newblock {\em Phys. Scripta}, 82(4):045013, 2010.

\bibitem{DHS15}
M.~Darroodi, H.~Hassanabadi, and N.~Salehi.
\newblock {\em Eur. Phys. J. A}, 51(6):69, 2015.

\bibitem{CMB04}
L.~Chetouani, M.~Merad, T.~Boudjedaa, and A.~Lecheheb.
\newblock {\em Int. J. Theor. Phys.}, 43(4):1147, 2004.

\bibitem{BBo05}
B.~Boutabia-Chéraitia and T.~Boudjedaa.
\newblock {\em Phys. Lett. A}, 338(2):97, 2005.

\bibitem{Bou08}
A.~Boumali.
\newblock {\em J. Math. Phys.}, 49(2):022302, 2008. Erratum: \emph{J. Math.
  Phys.}, 54(9):099902, 2013.

\bibitem{CTC10}
Y.~Chargui, A.~Trabelsi, and L.~Chetouani.
\newblock {\em Phys. Lett. A}, 374(29):2907, 2010.

\bibitem{CCC10}
T~R Cardoso, L~B Castro, and A~S de~Castro.
\newblock {\em J. Phys. A: Math. Theor.}, 43(5):055306, 2010.

\bibitem{Cas10}
A.~S. de~Castro.
\newblock {\em J. Math. Phys.}, 51(10):102302, 2010.

\bibitem{BBo12}
B.~Boutabia-Chéraitia and T.~Boudjedaa.
\newblock {\em J. Geom. Phys.}, 62(10):2038, 2012.

\bibitem{CTr13}
Y.~Chargui and A.~Trabelsi.
\newblock {\em Phys. Scripta}, 87(6):065003, 2013.

\bibitem{Bou15}
A.~Boumali.
\newblock {\em Z. Naturforsch. A}, 70(10):867, 2015.

\bibitem{dOl16}
L.~P. de~Oliveira.
\newblock {\em Ann. Phys.}, 372(Supplement C):320, 2016.

\bibitem{FMB17}
M.~Falek, M.~Merad, and T.~Birkandan.
\newblock {\em J. Math. Phys.}, 58(2):023501, 2017.

\bibitem{Tok66}
Z.~Tokuoka.
\newblock {\em Nucl. Phys.}, 78(3):681, 1966.

\bibitem{FNS73}
E.~Fischbach, M.~Martin Nieto, and C.~K. Scott.
\newblock {\em J. Math. Phys.}, 14(12):1760, 1973.

\bibitem{Fuj53}
Fujiwara I.
\newblock {\em Prog. Theor. Phys.}, 10:589, 1953.

\end{thebibliography}

\end{document}